\documentclass[a4paper]{article}
\usepackage{INTERSPEECH2020}
\usepackage{enumitem}
 \usepackage{multirow}
\title{Alzheimer's Dementia Recognition through Spontaneous Speech: \\The ADReSS Challenge}
\name{Saturnino Luz$^1$, Fasih Haider$^1$, Sofia de la Fuente$^1$, Davida Fromm$^2$,  Brian MacWhinney$^2$}
\address{
  $^1$Usher Institute, Edinburgh Medical School, The University of
  Edinburgh, UK\\
  $^2$Department of Psychology, Carnegie Mellon University, USA}
\email{\{S.Luz, fasih.haider, sofia.delafuente\}@ed.ac.uk, \{fromm, macw\}@andrew.cmu.edu}

\begin{document}

\maketitle
\begin{abstract}  
  The ADReSS Challenge at INTERSPEECH 2020 defines a shared task
  through which different approaches to the automated recognition of
  Alzheimer's dementia based on spontaneous speech can be compared.
  ADReSS provides researchers with a benchmark speech dataset which
  has been acoustically pre-processed and balanced in terms of age and
  gender, defining two cognitive assessment tasks, namely: the
  Alzheimer's speech classification task and the neuropsychological
  score regression task. In the Alzheimer's speech classification
  task, ADReSS challenge participants create models for classifying
  speech as dementia or healthy control speech. In the the
  neuropsychological score regression task, participants create models
  to predict mini-mental state examination scores. This paper
  describes the ADReSS Challenge in detail and presents a baseline for
  both tasks, including feature extraction procedures and results for
  classification and regression models.  ADReSS aims to provide the
  speech and language Alzheimer's research community with a platform
  for comprehensive methodological comparisons. This will hopefully
  contribute to addressing the lack of standardisation that currently
  affects the field and shed light on avenues for future research and
  clinical applicability.
\end{abstract}
\vspace{0.3cm}

\section{Introduction}
Alzheimer's Disease (AD) is a neurodegenerative disease that entails a
long-term and usually gradual decrease of cognitive functioning
\cite{AmericanPsychiatricAssociation2000}. It is also the most common
underlying cause for dementia. The main risk factor for AD is age, and
therefore its greatest incidence is amongst the elderly. Given the
current demographics in the Western world, where the population aged
65 years or more has been predicted to triple between years 2000 and
2050 \cite{WorldHealthOrganization2013}, institutions are investing
considerably on dementia prevention, early detection and disease
management. There is a need for cost-effective and scalable methods
that are able to identify the most subtle forms of AD, from the
preclinical stage of Subjective Cognitive Decline (SCI), to more
severe conditions like Mild Cognitive Impairment (MCI) and Alzheimer's
Dementia (AD) itself.

Whilst memory is often considered the main symptom of AD, language is also deemed as a valuable source of clinical information. Furthermore, the ubiquity of speech has led to a number of studies investigating speech and language features for the detection of AD, such as \cite{fraser2016linguistic, LUZ18.5, mirheidari2018detecting, haider2019assessment} to cite some examples. Although these studies propose various signal processing and machine learning methods for this task, the field still lacks balanced and standardised datasets on which these different approaches could be systematically compared.

Consequently, the main objective of the ADReSS Challenge of
INTERSPEECH 2020 is to define a shared task through which different
approaches to AD detection, based on spontaneous speech, could be
compared. This aims to address one of the main problems of this active
research field, the lack of standardisation, which hinders its
translation into clinical practice.  The ADReSS Challenge will therefore:
1) target a difficult automatic prediction problem of societal and medical relevance, namely, the detection of cognitive impairment and Alzheimer's Dementia (AD);
2)  to provide a forum for those different research groups to test
their existing methods (or develop novel approaches) on a new shared
standardized dataset;
3) mitigate common biases often overlooked in evaluations of AD detection methods, including repeated occurrences of speech from the same participant (common in longitudinal datasets), variations in audio quality, and imbalances of gender and age distribution; and
4) focus on AD recognition using spontaneous speech, rather than speech samples that are collected under laboratory conditions.

To the best of our knowledge, this will be the first such shared-task
focused on AD. Unlike some tests performed in clinical settings, where
short speech samples are collected under controlled conditions, this
task focuses on AD recognition using spontaneous speech. While a number
of researchers have proposed speech processing and natural language
processing approaches to AD recognition through speech, their studies
have used different, often unbalanced and acoustically varied datasets, 
consequently hindering reproducibility, replicability, and
comparability of approaches. The ADReSS Challenge will provide a forum
for those different research groups to test their existing methods (or
develop novel approaches) on a shared dataset which consists of a
statistically balanced, acoustically enhanced set of recordings of
spontaneous speech sessions along with segmentation and detailed
timestamped transcriptions. The use of spontaneous speech also sets
the ADReSS Challenge apart from tests performed in clinical settings
where short speech samples are collected under controlled conditions
which are arguably less suitable for the development of large-scale
monitoring technology than spontaneous speech
\cite{luz2017longitudinal}.

As data scarcity and heterogeneity have hindered research into the
relationship between speech and AD, the ADReSS Challenge provides
researchers with the very first available benchmark,
acoustically pre-processed and balanced in terms of age and gender.
ADReSS defines two different prediction tasks:
%\begin{itemize}
%\item
(a) the \textit{AD recognition task}, which requires researchers to
model participants' speech data to perform a binary classification of
speech samples into AD and non-AD classes; and
%\item
(b) the \textit{MMSE prediction task}, which requires researchers to
create regression models of the participants' speech in order to
predict their scores in the Mini-Mental State Examination (MMSE).
%\end{itemize}

This paper presents baselines for both tasks, including feature
extraction procedures and initial results for a classification and a
regression model.

\section{ADReSS Challenge Dataset}
\label{sec:adress-chall-dataset}

A dataset has been created for this challenge which is matched for age
and gender, as shown in Table~\ref{training} and Table~\ref{test}, so
as to minimise risk of bias in the prediction tasks. The data consists
of speech recordings and transcripts of spoken picture descriptions
elicited from participants through the Cookie Theft picture from the
Boston Diagnostic Aphasia Exam
\cite{Becker1994,bib:GoodglassKaplanBarresi01b}.  Transcripts were
annotated using the CHAT coding system
\cite{macwhinney2014childes}. The recorded speech has been segmented
for voice activity using a simple voice activity detection algorithm
based on signal energy threshold.  We set the log energy threshold
parameter to 65~dB with a maximum duration of 10 seconds per speech
segment. The segmented dataset contains 1,955 speech segments from 78
non-AD subjects and 2,122 speech segments from 78 AD subjects. The
average number of speech segments produced by each participant was
24.86 (standard deviation $sd=12.84$). Recordings were acoustically
enhanced with stationary noise removal and audio volume normalisation
was applied across all speech segments to control for variation caused
by recording conditions such as microphone placement.

% source('~/lib/projects/DTP/HRI-Dementia/R/dementiabank.R')
% ad <- loadADReSSdata()
% tapply(ad$train$mmse, list(ad$train$age.interval, ad$train$ad), mean, na.rm=T)
% tapply(ad$train$mmse, ad$train$ad, mean, na.rm=T)
% tapply(ad$train$mmse, ad$train$ad, sd, na.rm=T)
\begin{table}[h]
\centering
\caption {ADReSS Training Set: Basic characteristics of the patients in each group (M=male and F=female).}\vspace{-1ex}
\label{training}\footnotesize
\begin{tabular}{lrrrrrr}
\hline
 &  \multicolumn{3}{c}{AD}  &\multicolumn{3}{c}{non-AD}  \\
Age          & M & F & MMSE (sd) &  M & F & MMSE (sd)\\ \hline 
$\left[50, 55\right)$&  1 &  0 & 30.0 (n/a)&  1 &  0 & 29.0 (n/a)\\
$\left[55, 60\right)$&  5 &  4 & 16.3 (4.9)&  5 &  4 & 29.0 (1.3)\\ 
$\left[60, 65\right)$&  3 &  6 & 18.3 (6.1)&  3 &  6 & 29.3 (1.3)\\ 
$\left[65, 70\right)$&  6 & 10 & 16.9 (5.8)&  6 & 10 & 29.1 (0.9)\\ 
$\left[70, 75\right)$&  6 &  8 & 15.8 (4.5)&  6 &  8 & 29.1 (0.8)\\ 
$\left[75, 80\right)$&  3 &  2 & 17.2 (5.4)&  3 &  2 & 28.8 (0.4)\\ \hline 
  Total		     & 24 & 30 & 17.0 (5.5)& 24 &  30& 29.1 (1.0)\\ \hline
\end{tabular}
\end{table}

\begin{table}[h]
\centering
\caption {Characteristics of the ADReSS test set.}\vspace{-1ex}
\label{test}\footnotesize
\begin{tabular}{lrrrrrr}
\hline
 &  \multicolumn{3}{c}{AD}  &\multicolumn{3}{c}{non-AD}  \\
Age          & M & F & MMSE (sd) &  M & F & MMSE (sd)\\ \hline 
$\left[50, 55\right)$& 1 & 0 & 23.0 (n.a)& 1 & 0& 28.0 (n.a)  \\
$\left[55, 60\right)$& 2 & 2 & 18.7 (1.0)& 2 & 2& 28.5 (1.2)  \\ 
$\left[60, 65\right)$& 1 & 3 & 14.7 (3.7)& 1 & 3& 28.7 (0.9)   \\ 
$\left[65, 70\right)$& 3 & 4 & 23.2 (4.0)& 3 & 4& 29.4 (0.7)  \\ 
$\left[70, 75\right)$& 3 & 3 & 17.3 (6.9)& 3 & 3& 28.0 (2.4)  \\ 
$\left[75, 80\right)$& 1 & 1 & 21.5 (6.3)& 1 & 1& 30.0 (0.0)   \\ \hline 
  Total		&     11 &13 & 19.5 (5.3)&11 &13& 28.8 (1.5)  \\ \hline
\end{tabular}
\end{table}

\section{Acoustic and Linguistic Features}
\label{sec:acoust-feat-extr}

Acoustic feature extraction
was performed on the speech segments using the openSMILE v2.1 toolkit
which is an open-source software suite for automatic extraction of
features from speech, widely used for emotion and affect recognition
in speech \cite{eyben2010opensmile}, and with in-house software
\cite{bib:Luz18vocaldia}. As the purpose of this paper is to describe
the prediction tasks and set simple baselines that can be attained
without extensive optimisation, we did not perform any feature set
reduction procedures.  The following is a brief description of the
acoustic feature sets used
in the experiments described in this paper:

\textit{emobase:} This feature set contains the mel-frequency
cepstral coefficients (MFCC) voice quality, fundamental frequency
(F0), F0 envelope, line spectral pairs (LSP) and intensity features with their first and
second order derivatives. Several statistical functions are applied to
these features, resulting in a total of 988 features for every speech
segment \cite{eyben2010opensmile}.

\textit{ComParE:} The \textit{ComParE 2013} \cite{eyben2013recent}
feature set includes energy, spectral, MFCC, and voicing related
low-level descriptors (LLDs). LLDs include logarithmic
harmonic-to-noise ratio, voice quality features, Viterbi smoothing for
F0, spectral harmonicity and psychoacoustic spectral
sharpness. Statistical functionals are also computed, bringing the
total to 6,373 features.

\textit{eGeMAPS:} The \textit{eGeMAPS} \cite{eyben2016geneva} feature
set resulted from an attempt to reduce the somewhat unwieldy feature
sets above to a basic set of acoustic features based on their
potential to detect physiological changes in voice production, as well
as theoretical significance and proven usefulness in previous studies
\cite{Eyben2016}. It contains the F0 semitone, loudness,
spectral flux, MFCC, jitter, shimmer, F1, F2, F3, alpha ratio,
Hammarberg index and slope V0 features, as well as their most common
statistical functionals, for a total of 88 features per speech
segment.

\textit{MRCG functionals:} Multi-resolution Cochleagram features (MRCGs) were proposed by Chen et
al. \cite{chen2014feature} and have since been used in speech related
applications such as voice activity detection \cite{kim2018voice},
speech separation \cite{chen2014feature}, and more recently for
attitude recognition \cite{bib:HaiderLuz19atrusm}. MRCG features are
based on cochleagrams \cite{wang2005ideal}. A cochleagram is generated
by applying the gammatone filter to the audio signal, decomposing it
in the frequency domain so as to mimic the human auditory
filters. MRCG uses the time-frequency representation to encode the
multi-resolution power distribution of the audio signal. Four
cochleagram features were generated at different levels of
resolution. The high resolution level encodes local information while
the remaining three lower resolution levels capture spectrotemporal
information.  A total of 768 features were extracted from each frame:
256 MRCG features (frame length of 20 ms and frame shift of 10 ms),
along with 256 $\Delta$ MRCG and 256 $\Delta\Delta$ MRCG
features. The statistical functionals (mean,
standard deviation, minimum, maximum, range, mode, median, skewness
and kurtosis) were applied on the 768 MRCG features for a total of 6,912 features.

\textit{Minimal:} this feature set consists of basic statistics (mean,
standard deviation, median, minimum and maximum) of the duration of
vocalisations and pauses and speech rate, and a vocalisation count,
similarly to \cite{luz2017longitudinal}.

In sum, we extracted 88 eGeMAPS, 988 emobase, 6,373 ComParE, 6,912
MRCG, and 13 minimal features from 4,077 speech segments. Excepting
the minimal feature set, Pearson's correlation test was performed to
remove acoustic features that were significantly correlated with
duration (when $|R| > 0.2$). Hence, 72 eGeMAPS, 599 emobase, 3,056
ComParE, and 3,253 MRCG features were not correlated with the duration
of the speech chunks, and were therefore selected for the machine
learning experiments.  Examples of features from the ComParE feature
set by the above described procedure include L1-norms of segment
length functionals smoothed by a moving average filter (including
their means, maxima and standard deviations),
%% audspec\-\_lengthL1norm\-\_sma\-\_meanSegLen,
%% audspec\-\_lengthL1norm\-\_sma\-\_maxSegLen, and
%% audspec\-\_lengthL1norm\-\_sma\-\_segLenStddev),
and the relative spectral
transform applied to auditory spectrum (RASTA) functionals (including
the percentage of time the signal is above 25\%, 50\% and 75\% of
range plus minimum). 
%% (aud\-spec\-Rasta\-\_length\-L1norm\-\_sma\-\_upleveltime25,
%% aud\-spec\-Rasta\-\_length\-L1norm\-\_sma\-\_upleveltime50, and
%% aud\-spec\-Rasta\-\_length\-L1norm\-\_sma\-\_upleveltime75).

In addition, we used the EVAL command in the CLAN program
\cite{bib:MacWhinneyt2017} to compute a basic set of 34 language
outcome measures (e.g., duration, total utterances, MLU, type-token
ratio, open-closed class word ratio, percentages of 9 parts of speech)
on the CHAT transcripts.

\section{AD classification task}

The AD classification task consists of creating a binary
classification models to distinguish between AD and non-AD patient
speech. These models may use speech data, transcribed speech, or
both. Any methodological approach may be taken, but participants will
work with the same dataset. The evaluation metric for this task are
$\operatorname {Accuracy} =  {\frac { TN + TP }{N} }$, precision 
${\pi} =  { \frac { TP }{TP + FP} }$, recall $\operatorname {\rho} =
{ \frac { TP }{TP + FN} }$, and $\operatorname {F_1} =  { 2 \frac { \pi \times \rho
  }{\pi + \rho} }$, where  N is the number of patients, TP, FP and FN
are the number of true positives, false positives and false negatives,
respectively. 

%\subsection{Baseline classification}

We performed our baseline classification experiments using five
different methods, namely linear discriminant analysis (LDA),
decision trees (DT, with leaf size of 20 and the CART algorithm),
nearest neighbour (1NN, for KNN with K=1), 
random forests (RF, with 50 trees and a leaf size of 20) and support
vector machines (SVM, with a linear kernel with box constraint of 0.1,
and sequential minimal optimisation solver).  The classification
methods were implemented in MATLAB \cite{bib:MATLAB19} using the
statistics and machine learning toolbox. A leave-one-subject-out
(LOSO) cross-validation setting was adopted, where the training data
do not contain any information from validation subjects.

\begin{figure*}[htb]
	\centering
	\centerline{\includegraphics[width=.65\linewidth]{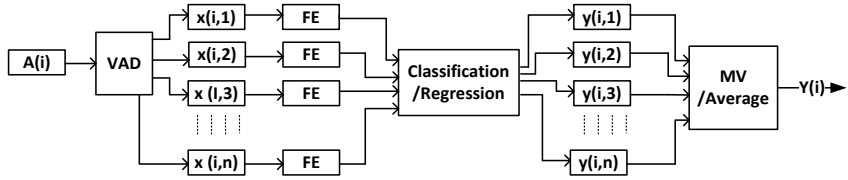}}
	\caption{System Architecture: $A(i)$, the recording of
          is segmented using voice activity
          detection (VAD) into $n$ segments $x(i,n)$. Acoustic feature
          extraction (FE) is performed at segment level. The output of
          classification or regression for the $n^{th}$ segment of the
          $i^{th}$ recording is denoted $y(i,n)$.  MV outputs the
          majority voting for classification, and Average the 
          mean regression score.}
    \label{fig:system}  
\end{figure*}

%\subsection{Classification experiments}

Two-step classification experiments were conducted to detect cognitive
impairment due to AD (as shown in Figure~\ref{fig:system}).  This
consisted of segment-level (SL) classification, where classifiers were
trained and tested to predict
whether a speech segment was uttered by a non-AD or AD patient, and
majority vote (MV) classification, which assigned each subject a class
label based on the majority labels of SL classification.

\subsection{Results}

The classification accuracy is shown in Tables~\ref{results-MV-val} and
\ref{results-MV-test} for LOSO and test settings respectively.  These
results show that the 1NN (0.574) provides the best accuracy for acoustic features using
ComParE set for AD detection, with accuracy above the chance level
of 0.50. 
From the results shown in Table~\ref{results-MV-val}, we note that
even though 1NN provides the best result (0.574), DT and LDA also
exhibit promising performance, being in fact more stable across all
feature sets than the other classifiers (the best average accuracy of
0.559 for LDA and 0.570 for DT).  We also note that Minimal, ComParE
and linguistic also exhibit promising performance, being in fact more stable
across all classifiers than the other features (the best average
accuracy of 0.552 for Minimal, 0.541 for Compare and 0.713 for
linguistic). Based on these findings we have selected the LDA model
trained using ComParE as our baseline model for acoustic features.

Table~\ref{results-MV-test} shows that 1NN provides less accurate
results on the test set than in LOSO cross validation. However, the
results of LDA (0.625) and DT (0.625) improve on the test data for
acoustic features. The linguistic features provide an accuracy of
0.75, which is better than automatically extracted acoustic features
though it relies on manual transcription.
%For further insight,
%the confusion matrices for acoustics (i.e. \textit{ComParE}) and
%linguistic features are shown in Figure~\ref{fig:task1-test}.  Hence
The challenge baseline accuracy for the classification task are therefore 0.625
for acoustic features and 0.75 for linguistic features.  The
precision, recall and F1 Score are reported in
Table~\ref{tab:baseline}.

\begin{table}[!h]
\centering
\caption{AD classification accuracy on LOSO cross validation.}
	\vspace{-2mm}
\resizebox{\columnwidth}{!}{%
\begin{tabular}{llllll|l}
\hline
  Features & LDA & DT & 1NN  & SVM  & RF  & mean \\ \hline
  emobase  & 0.500 & 0.519 & 0.398  & 0.491  & 0.472   & 0.476\\ 
  ComParE  & \textbf{0.565} & 0.528 & 0.574 & 0.528  & 0.509  & \textbf{0.541}\\ 
  eGeMAPS  & 0.482 &  0.500 & 0.380  & 0.333  & 0.482  & 0.435\\ 
  MRCG     & 0.519 & 0.500 & 0.482 & 0.528  & 0.509  & 0.507\\ 
  Minimal  & 0.519 & 0.667 & 0.426 & 0.565 & 0.583 &  0.552 \\ 
  linguistic  &  \textbf{0.768} & 0.704 & 0.740 & 0.602 & 0.750  &  \textbf{0.713} \\ \hline
  mean     &    \textbf{0.559}  &  \textbf{0.570}  &  0.500  &  0.508 &   0.551  &   -- \\ \hline
\end{tabular}
}
\label{results-MV-val}
\end{table}

% \begin{figure}[htb]
% 	\centering
% 	\centerline{\includegraphics[width=.8\linewidth]{acoustic.eps}}
% 		\vspace{-2mm}
% 	\caption{Confusion matrix for acoustic features.}
%     \label{fig:task1-val}  
% \end{figure}

\begin{table}[!h]
\centering
\caption{AD classification accuracy on test set.}
	\vspace{-2mm}
\resizebox{\columnwidth}{!}{%
\begin{tabular}{llllll|l}
\hline
  Features & LDA & DT & 1NN  & SVM  & RF & mean \\ \hline
  emobase  & 0.542   &   0.688  &    0.604  &    0.500   &   0.729  & 0.613 \\ 
  ComParE  &  \textbf{0.625}  &  0.625  &  0.458 &   0.500  &  0.542   & 0.550\\ 
  eGeMAPS  &  0.583 &   0.542  &  0.688 &   0.563  &  0.604  & 0.596\\ 
  MRCG     &  0.542 &   0.563  &  0.417 &   0.521 &   0.542  & 0.517\\
  Minimal  & 0.604 & 0.562 & 0.604 & 0.667 & 0.583 &  0.604 \\ 
  linguistic  &  \textbf{0.750}  &   0.625  &   0.667  &   0.792 &    0.750   &  0.717\\ \hline
  mean     &    0.608  &    0.601   &   0.573   &   0.590   &   0.625  & --\\ \hline
\end{tabular}
}

\label{results-MV-test}
\end{table}

% \begin{figure}[htb]
% 	\centering
% 	\centerline{\includegraphics[width=0.8\linewidth]{ling.eps}}
% 	\vspace{-2mm}
% 	\caption{Confusion matrix for linguistic features. }
%     \label{fig:task1-test}  
%     \vspace{-5mm}
% \end{figure}

\begin{table}[!h]
\centering
\caption{Baseline results of AD classification task using the LDA classifier with \textit{acoustic and 
linguistic} features.}
	\vspace{-2mm}
	\resizebox{\columnwidth}{!}{%
\begin{tabular}{lllllll}
\hline
  & class &  Precision & Recall & F1 Score & Accuracy\\  \hline
\multirow{2}{*}{$LOSO_{Acous}$} & non-AD &  0.56 & 0.61 & 0.58 &  \multirow{2}{*}{0.56}\\
                  & AD &  0.57 & 0.52 & 0.54  & \\ \hline  
\multirow{2}{*}{$TEST_{Acous}$} & non-AD &  0.67 & 0.50 & 0.57  &  \multirow{2}{*}{0.62}\\
  & AD &  0.60 & 0.75 & 0.67 & \\ \hline\hline
  
  \multirow{2}{*}{$LOSO_{ling}$} & non-AD & 0.76  & 0.78 & 0.77 &  \multirow{2}{*}{0.77}\\
                  & AD & 0.77  & 0.76 &  0.77 & \\ \hline  
\multirow{2}{*}{$TEST_{ling}$} & non-AD & 0.70  & 0.87 & 0.78  &  \multirow{2}{*}{0.75}\\
  & AD &  0.83 &  0.62 & 0.71 & \\
\hline                               
\end{tabular}
}
\vspace{-5mm}
\label{tab:baseline}
\end{table}

\section{MMSE prediction task}
\label{sec:mmse-prediction-task}

The MMSE prediction task consists of generating a regression model for
prediction of MMSE scores of individual participants from the AD and
non-AD groups. Unlike classification, MMSE prediction is relatively
uncommon in the literature, despite MMSE scores often being
available. While models may use speech (acoustic) or linguistic data
individually or in combination, the baseline described here report
results of acoustic and linguistic models built separately.

\subsection{Baseline regression}

We performed our baseline regression experiments using five different
methods, namely decision trees (DT, with leaf size of 20 and CART algorithm), linear
regression (LR), gaussian process regression (GPR, with a squared
exponential kernel), least-squares boosting (LSBoost, which contains
the results of boosting 100 regression trees) and support vector
machines (SVM, with a radial basis function kernel with box constraint
of 0.1, and sequential minimal optimisation solver).  The regression
methods are implemented in MATLAB \cite{bib:MATLAB19} using the
statistics and machine learning toolbox.  As with classification, the
regression experiments were conducted in two steps for acoustic features
(Figure~\ref{fig:system}), with SL regression followed by averaging of
predicted MMSE values.

\subsection{Results}
The regression results are reported as root mean squared error (RMSE)
scores in Tables~\ref{results-val-mmse} and \ref{results-test-mmse}
for LOSOCV and test data.  These results show that DT (7.28) provides
the best RMSE using MRCG features for MMSE prediction with
$r= -0.759$, being more stable across all acoustic feature sets than
the other classifiers (the best average RMSE of 6.86 for DT). We also
note that Minimal and eGeMaPs also exhibit promising performance, with
RMSE of 7.46 and 8.02 respectively across models.  Based on this, the
DT model trained using the MRCG feature was chosen as the baseline
model for the regression task for acoustic features. For linguistic
features, we selected the DT model as baseline with RMSE of 4.38 ($r = 0.792$).

Table~\ref{results-test-mmse} shows the results of regression methods
on test data. The baseline model (DT with MRCG features) provides an
RMSE of 6.14 ($r = 0.22$) in the test setting. Hence the challenge
baseline accuracy for this task is 6.14 for acoustic features. The
linguistic feature model provides an RMSE of 5.20 ($r=
0.57$), which therefore corresponds to the ADReSS challenge baseline
accuracy for linguistic features in this task.

\begin{table}[!h]
\centering
\caption{MMSE prediction LOSO cross Validation results. the chance level is RMSE of 7.18} 
	\vspace{-2mm}
\resizebox{\columnwidth}{!}{%
\begin{tabular}{llllll|l}
%\cline{2-6}
\hline
  Features &  Linear & DT & GP  & SVM  & LSBoost  & mean\\  \hline
  emobase & 7.44  & 7.29 & 7.71 & 7.71 & 8.33 & 7.70  \\ 
  ComParE & 15.69  & 7.29 & 7.67 & 7.63 & 7.84 & 9.22\\ 
  eGeMAPS & 8.08 & 7.31 & 7.72 & 8.55 & 8.68 & 8.07  \\
  MRCG & 13.46 & \textbf{7.28, r= -0.76} & 7.65 & 7.50 & 8.02 & 8.78 \\
  Minimal & 7.39 & 7.60 & 7.18 & 8.01 & 7.14 &  7.46 \\
  Linguistic &  6.15  &   \textbf{4.38, r=0.79} &    7.92 &    6.34 &    7.44 & 6.45   \\
  \hline
mean & 9.70  &  \textbf{6.86}  &  7.64  &  7.62  &  7.91   & --\\ \hline
\end{tabular}
}
\label{results-val-mmse}
\end{table}
	\vspace{-5mm}
\begin{table}[!h]
\centering
\caption{MMSE prediction test results.}
%%6.06
	\vspace{-2mm}
\resizebox{\columnwidth}{!}{%
\begin{tabular}{llllll|l}
%\cline{2-6}
\hline
  Features & Linear & DT & GP & SVM & LSBoost & mean\\  \hline
  emobase  & 6.80 & 6.78 & 6.36 & 6.18 & 6.73 & 6.57\\ 
  ComParE  & 6.47 & 6.52 & 6.33 & 6.19 & 6.72 & 6.45  \\ 
  eGeMAPS  & 6.90 & 5.99 & 6.28 & 6.12 & 6.41 & 6.34 \\
  MRCG &     6.70 & \textbf{6.14, r=0.22} & 6.33 & 6.20 & 6.31 &  6.33\\
  Minimal  & 6.29 & 6.84 & 6.58 & 6.19 & 7.71 & 6.72  \\
  Linguistic &    4.78  &  \textbf{5.20, r= 0.57}  &  5.54  &  6.24  &  6.62  & 5.68  \\
  \hline
  mean &     6.32  &  \textbf{6.25} &  6.24  &  6.19  &  6.75  &-- \\ \hline
\end{tabular}
}
\label{results-test-mmse}
\end{table}
\vspace{-5mm}
\section{Discussion}

These results of the classification baseline are comparable to those
attained by models based on speech recordings available from
spontaneous speech samples in DementiaBank's Pitt corpus
\cite{Becker1994}, which is widely used. Accuracy scores of 81.92\%,
80\% and 79\% and 64\% have been reported in the literature
\cite{fraser2016linguistic,yancheva2016vector,hernandez2018computer,luz2017longitudinal}. Although
these studies report higher accuracy than ours, all of those (except
\cite{luz2017longitudinal}) include information from the manual
transcripts, and were conducted on an unbalanced dataset (in terms of
age, gender and number of subjects in the AD and non-AD classes). It
is also worth noting that accuracy for the best performing of these
models drops to 58.5\% when feature selection is not performed on
their original set of 370 linguistic and acoustic features
\cite{fraser2016linguistic}.  The performance of a model without the
information from transcripts, that is, relying only on acoustic
features as we do, is only reported in \cite{luz2017longitudinal}
(64\%) and \cite{hernandez2018computer}, where its SVM model drops to
an average accuracy of 62\%.  It is also noted that previous studies
do not evaluate their methods in a complete subject-independent
setting (i.e. they consider multiple sessions for a subject and
classify a session instead of a subject). This could lead to
overfitting, as the model might learn speaker dependent features from
a session and then, based on those features, classify the next session
of the same speaker.

One strength of our method is its speaker independent nature.
Ambrosini et al. reported an accuracy of 80\% while using acoustic
(pitch, unvoiced duration, shimmer, pause duration, speech rate), age
and educational level features for cognitive decline detection using
an Italian dataset of an episodic story telling setting
\cite{ambrosini2019automatic}. However, this dataset is less easily
comparable to ours, as it is elicited differently, and is not age and
gender balanced.

Yancheva and colleagues \cite{yancheva2015using} predicted MMSE scores
with speech-related features using the full DementiaBank Pitt dataset,
which is not balanced and includes longitudinal observations. Their
model yielded a mean absolute error (MAE) of 3.83 in predicting
MMSE. However, they employed lexicosyntactic and semantic features
derived from manual transcription, rather than automatically extracted
acoustic features as we used in our analysis. In
\cite{yancheva2015using}, those linguistic features were the main
features selected from a group of 477, with acoustic features
typically not being among the most relevant. Therefore no quantitative
results were reported for acoustic features.

\section{Conclusions}

This paper described the ADReSS challenge, and set simple baselines
for its tasks, demonstrating the relevance of acoustic and linguistic
features of spontaneous speech for cognitive impairment detection in
the context of Alzheimer's Disease diagnosis and MMSE
prediction. Machine learning methods operating on automatically
extracted voice features provide a baseline accuracy of up to 62.5\%
on the AD classification task, while linguistic features extracted
from manually produced transcripts yielded 76.85\% accuracy on the
same task. These results are well above the chance level of 50\%.  A
baseline RMSE of 6.14 and 5.21 for acoustic and linguistic features
respectively on test has been established for the MMSE regression
task. It is reasonable to expect that the ADReSS Challenge's
participants will attain better accuracy scores by employing further
pre-processing, feature set reduction, and more complex models than
the ones employed in this paper.  By bringing the research community
together in order to work on a shared task on the same dataset, ADReSS
intends to make comprehensive methodological comparisons. This will
hopefully highlight research caveats and shed light on avenues for
clinical applicability and future research directions.

\section{Acknowledgements}
This research is funded by the European Union's Horizon 2020 research
programme, under grant agreement 769661, SAAM project.
SdlFG is supported by the Medical Research Council.

\bibliographystyle{IEEEtran}
\bibliography{IEEEabrv,ADReSS}

% \begin{thebibliography}{9}
% \bibitem[1]{Davis80-COP}
%   S.\ B.\ Davis and P.\ Mermelstein,
%   ``Comparison of parametric representation for monosyllabic word recognition in continuously spoken sentences,''
%   \textit{IEEE Transactions on Acoustics, Speech and Signal Processing}, vol.~28, no.~4, pp.~357--366, 1980.
% \bibitem[2]{Rabiner89-ATO}
%   L.\ R.\ Rabiner,
%   ``A tutorial on hidden Markov models and selected applications in speech recognition,''
%   \textit{Proceedings of the IEEE}, vol.~77, no.~2, pp.~257-286, 1989.
% \bibitem[3]{Hastie09-TEO}
%   T.\ Hastie, R.\ Tibshirani, and J.\ Friedman,
%   \textit{The Elements of Statistical Learning -- Data Mining, Inference, and Prediction}.
%   New York: Springer, 2009.
% \bibitem[4]{YourName17-XXX}
%   F.\ Lastname1, F.\ Lastname2, and F.\ Lastname3,
%   ``Title of your INTERSPEECH 2020 publication,''
%   in \textit{Interspeech 2020 -- 20\textsuperscript{th} Annual Conference of the International Speech Communication Association, September 15-19, Graz, Austria, Proceedings, Proceedings}, 2020, pp.~100--104.
% \end{thebibliography}

\end{document}